\def\fr#1#2{\hbox{${#1\over #2}$}}            
                            \def\pd{\partial}          
\def\ort{\perp}                               \def\Ra{\Rightarrow}

\def\nul{{(0)}}          \def\one{{(1)}}
\def\m{\mu}              \def\n{\nu}              \def\k{\kappa}
           \def\g{\gamma}           \def\d{\delta}
\def\S{\Sigma}           \def\s{\sigma}           \def\t{\tau}  
\def\a{\alpha}           \def\b{\beta}            \def\th{\theta}
       \def\D{\Delta}           \def\l{\lambda}
\def\ve{\varepsilon}     \def\p{\pi}               
\def\L{\Lambda}
\def\cL{{\cal L}}        \def\cH{{\cal H}}       
\def\hp{{\hat \pi}}
\def\bi{{\bar i}}        \def\bk{{\bar k}}        \def\bl{{\bar l}}
\def\bm{{\bar m}}        \def\bn{{\bar n}}        \def\bj{{\bar j}}
\def\bT{{\bar T}{}}            
\def\bcL{{\bar{\cal L}}{}}       \def\tphi{\tilde\phi}
\def\lra{\leftrightarrow}
\def\tgr{GR$_{\parallel}$}

\documentstyle[preprint,aps,eqsecnum]{revtex}
\def\nn{\nonumber}                    \def\subsub#1{{\bf #1}}
\def\be{\begin{equation}}             \def\ee{\end{equation}}
\def\bea{\begin{eqnarray} }           \def\eea{\end{eqnarray} }
\def\ba#1{\begin{array}{#1}}          \def\ea{\end{array}}
\def\lab#1{\label{#1}}                \def\eq#1{(\ref{#1})}
\def\bsubeq{\begin{mathletters}}      \def\esubeq{\end{mathletters}}
\def\bitem{\begin{itemize}}           \def\eitem{\end{itemize}}
\def\mb#1{\hbox{\boldmath $#1$}}

\begin{document}
\tighten
\title{Hamiltonian structure of the teleparallel formulation of GR}
\author{M. Blagojevi\'c\thanks{E-mail address:  mb@phy.bg.ac.yu}
        and I. A. Nikoli\'c}
\address{Institute of Physics, P.O.Box 57, 11001 Belgrade, Yugoslavia}
\date{February 01, 2000}
\maketitle
\begin{abstract}
We apply Dirac's Hamiltonian approach to study the canonical structure
of the teleparallel form of general relativity without matter fields.
It is shown, without any gauge fixing, that the Hamiltonian has the
generalized Dirac--ADM form, and constraints satisfy all the
consistency requirements. The set of constraints involves some extra
first class constraints, which are used to find additional gauge
symmetries and clarify the gauge structure of the theory. 
\end{abstract}

\section{Introduction} 

Among various attempts to overcome the problems of quantization and 
the existence of singular solutions in Einstein's general relativity
(GR), gauge theories of gravity are especially attractive, as they are
based on the concept of gauge symmetry which has been very successful
in the foundation of other fundamental interactions. The importance of
the Poincar\'e symmetry in particle physics leads one to consider the
Poincar\'e gauge theory (PGT) as a natural framework for description of
the gravitational phenomena \cite{1,2,3,4,5} (for more general
attempts, see \cite{6}).   

Basic gravitational variables in PGT are tetrad field $b^k{}_\m$ and 
Lorentz connection $A^{ij}{}_\m$, which are associated to the
translation and Lorentz subgroups of the Poincar\'e group,
respectively. These gauge fields are coupled to the energy--momentum 
and spin of matter fields, and their field strengths are geometrically
identified with the torsion and the curvature: 
$T^i{}_{\m\n}=\pd_\m b^i{_\n}+A^{i}{}_{s\m}b^s{_\n}-(\m\n)$,
$R^{ij}{}_{\m\n}=\pd_\m A^{ij}{_\n}+A^{i}{}_{s\m}A^{sj}{_\n}-(\m\n)$.
The spacetime of PGT turns out to be Riemann--Cartan space $U_4$,
equipped with metric and linear, metric compatible connection.   
Dynamical content of PGT is determined by the Lagrangian 
$\tilde\cL\equiv b(\cL_G +\cL_M)$, where the gravitational part $\cL_G$
is usually assumed to be at most quadratic in field strengths, and
$\cL_M$ describes minimally coupled matter fields. 

General geometric arena of PGT, the Riemann--Cartan space $U_4$, may be
{\it a priori\/} restricted by imposing certain conditions on the
curvature and the torsion.  Thus, Einstein's GR is defined in Riemann
space $V_4$, which is obtained from $U_4$ by the requirement of
vanishing torsion. Another interesting limit of PGT is 
{\it teleparallel\/} or {\it Weitzenb\"ock\/} geometry $T_4$, defined by
the requirement of vanishing curvature:       
\be
R^{ij}{}_{\m\n}(A)=0 \, .                                  \lab{1.1}
\ee
The vanishing of curvature means that parallel transport is path
independent, hence we have an absolute parallelism. 
The teleparallel geometry is, in a sense, complementary to Riemannian:
curvature vanishes, and torsion remains to characterize the parallel
transport. 

Of particular importance for the physical interpretation of the
teleparallel geometry is the fact that there is a one--parameter family
of teleparallel Lagrangians which is {\it empirically\/} equivalent to
GR \cite{5,7,8}. For the parameter value $B=1/2$ the Lagrangian of the
theory coincides, modulo a four--divergence, with the Einstein--Hilbert
Lagrangian, and defines the teleparallel form of GR, \tgr.  

The teleparallel description of gravity has been one of the most
promising alternatives to GR. However, analyzing this theory
Kopczy\'nsky \cite{9} found a hidden gauge symmetry, and concluded that
the torsion evolution is not completely determined by the field
equations. Assuming, then, that the torsion should be a measurable
physical quantity, he argued that this theory is internally
inconsistent.  Hayashi and Shirafuji \cite{10} tried to avoid this
problem by interpreting certain different torsion configurations as
physically equivalent, i.e. related to each other by a gauge
transformation, but the consistency of this idea in the interacting
theory seems to be questionable for non--scalar matter \cite{9,11}.
Various modifications of the one--parameter teleparallel theory are
proposed in order to avoid the above problems \cite{9,12,13}.  Trying
to re--examine the gauge structure of the one--parameter teleparallel
geometry Nester \cite{14} improved the arguments of  
Kopczy\'nsky \cite{9}; the predictability problem was stated more
precisely and bound to certain special solutions.  

Hecht et al. \cite{15} traced the appearance of non--physical modes of
torsion back to some symmetries which are necessarily present in the
(3+1) decomposition of spacetime. Using certain geometric arguments
they concluded that some components of the tetrad velocity are not
suited to represent dynamical degrees of freedom. In other words, these
velocities must not appear in the evolution equations, hence they
should appear at most linear in the Lagrangian. The choice of
parameters in the teleparallel Lagrangian that ensures this to happen
is just the one corresponding to \tgr.   

The teleparallel geometry possesses many salient features. Thus, Nester
\cite{16} succeeded in formulating a pure tensorial proof of the
positivity of total energy for Einstein's theory in terms of the
teleparallel geometry. He found that special gauge features of \tgr,
which are usually considered to be problematic, are quite beneficial
for this purpose. Mielke \cite{17} used the teleparallel geometry of
\tgr\ to give a transparent description of Ashtekar's complex
variables, while Andrade et al. \cite{18} formulated a
five--dimensional teleparallel equivalent of Kaluza--Klein theory. 
There are also some attempts to understand the role of torsion at the
quantum level \cite{19}.    

The purpose of this paper is to investigate the canonical structure of
\tgr\ using Dirac's Hamiltonian approach \cite{20}, as this is, in our
opinion, the best way to clarify both the nature of somewhat mysterious
extra gauge symmetries, and the question of consistency of \tgr.  
We shall find that a specific choice of coupling constants in
the teleparallel Lagrangian leads to the appearance of some additional 
first class constraints, and, consequently, to extra gauge symmetries,
which clarify the meaning of non--dynamical torsion components and give
us a complete picture of the gauge structure of \tgr.   

We remark here that Maluf \cite{21} tried to analyze some aspects of the
Hamiltonian structure of \tgr. However, his approach is based on
some unnecessary gauge fixing conditions, adopted at the level of
Lagrangian in order to simplify the calculations, so that many specific
gauge features of the theory remained effectively hidden.   

The layout of the paper is as follows. After recalling some basic
elements of the Lagrangian teleparallel formulation of GR in Sec. II,
we work out all the primary constraints and  construct the
corresponding Hamiltonian density in Sec. III. It is shown that a
specific choice of parameters in the Lagrangian leads to additional
primary constraints. Then, we study the consistency conditions in Sec.
IV, and derive the algebra of constraints in Sec. V. These results are
used in Sec. VI to construct extra gauge generators and clarify the
nature of the related gauge symmetries. Section VII is devoted to
concluding remarks, while some technical details are presented in the
Appendix.  

Our conventions are the same as in Ref. \cite{22}: the Latin indices
refer to the local Lorentz frame, whereas the Greek indices refer to
the coordinate frame; the first letters of both alphabets 
$(a,b,c,...;\a,\b,\g,...)$ run over 1,2,3, and the middle alphabet
letters $(i,j,k,...;\m,\n,\l,...)$ run over 0,1,2,3; 
$\eta_{ij}=$ diag $(+,-,-,-)$, $\ve^{0123}=+1$ and $\d=\d(x-x')$.

\section{The teleparallel formulation of GR} 

\subsub{Lagrangian.} Gravitational field in the framework of the
teleparallel geometry in PGT is described by the tetrad $b^k{_\m}$ and
Lorentz connection $A^{ij}{_\m}$, subject to the condition of vanishing
curvature \eq{1.1}. We shall consider here the gravitational dynamics
determined by a class of Lagrangians quadratic in the torsion
\cite{5,7,8}   
\bea
&&\tilde\cL=b\bigl(\cL_T +\l_{ij}{}^{\m\n}R^{ij}{}_{\m\n}
                        +\cL_M\bigr) \, , \nn \\
&&\cL_T=a\bigl(AT_{ijk}T^{ijk}+BT_{ijk}T^{jik}+CT_{k}T^{k}\bigr)
     \equiv \b_{ijk}(T)T^{ijk} \, ,                      \lab{2.1}
\eea 
where $\l_{ij}{}^{\m\n}$ are Lagrange multipliers introduced to ensure
the teleparallelism condition \eq{1.1} in the variational formalism,
$a=1/2\k$ ($\k=$ Einstein's gravitational constant), $T_k=T^m{}_{mk}$,
and $\cL_M$ is the Lagrangian of matter fields. The explicit form of
$\b_{ijk}$ is      
$$
\b_{ijk}=a\bigl(AT_{ijk}+BT_{[jik]}+C\eta_{i[j}T_{k}\bigr)\, .
$$

The parameters $A,B,C$ in the Lagrangian should be determined on
physical gro\-unds, so as to obtain a consistent theory which could
describe all the known gravitational experiments. If we require that
the theory \eq{2.1} gives the same results as GR in the li\-ne\-ar,
weak--field approximation, we can restrict our considerations to the
one--parameter family of Lagrangians, defined by the conditions 
\cite{5,7,8} 
\bitem
\item[$i)$] $\,2A+B+C=0\, ,\quad C=-1\, .$  
\eitem 
This family represents a viable gravitational theory for macroscopic,
spinless matter, empirically indistinguishable from GR. Von der Heyde
\cite{23} and Hehl \cite{5} have given certain theoretical arguments in
favor of the choice $B=0$. There is, however, another, particularly
interesting choice determined by the requirement 
\bitem 
\item[$ii)$] $\,2A-B=0$, 
\eitem 
It leads effectively to the Einstein--Hilbert Lagrangian
$\cL_{GR}=-abR(\D)$, defined in Riemann spacetime $V_4$ with
Levi--Civita connection $A=\D$, via the geometric identity \eq{A3}:
$$
bR(A)= bR(\D) +b\bigl( \fr{1}{4}T_{ijk}T^{ijk}
+\fr{1}{2}T_{ijk}T^{jik}-T_{k}T^{k}\bigr)-2\pd_\n(bT^{\n})\, .
$$
Indeed, in Weitzenb\"ock spacetime the above identity in conjunction
with the condition \eq{1.1} implies that the torsion Lagrangian in
\eq{2.1} is equivalent to the Einstein--Hilbert Lagrangian, up to a
four--divergence, provided that   
\be
A=\fr{1}{4}\, ,\qquad B=\fr{1}{2}\, ,\qquad C=-1  \, ,     \lab{2.2}
\ee
which coincides with the conditions $i)$ and $ii)$ given above.

The theory defined by equations \eq{2.1} and \eq{2.2} is called the
teleparallel formulation of GR (\tgr). Note that the equivalence with
GR holds certainly for scalar matter, while the gravitational couplings
to spinning matter fields in $T_4$ and $V_4$ are in general different.   

\subsub{Field equations.} By varying the Lagrangian \eq{2.1} with
respect to $b^i{_\m}, A^{ij}{_\m}$ and $\l_{ij}{}^{\m\n}$ we obtain the
gravitational field equations:
\bsubeq \lab{2.3}  
\bea
-&&4\nabla_\m\bigl(b\b_i{^{\m\n}}\bigr) -4b\b^{nm\n}T_{nmi}
                     +h_i{^\n}b\cL_T =\t^\n{_i}\, ,     \lab{2.3a}\\
-&&8b\b_{[ij]}{^\n}+4\nabla_\m\bigl( b\l_{ij}{^{\n\m}}\bigr)
                            =\s^\n{}_{ij} \, ,          \lab{2.3b}\\
 &&R^{ij}{}_{\m\n}=0 \, ,                               \lab{2.3c}
\eea
\esubeq  
where $\t^\n{_i}$ and $\s^\n{}_{ij}$ are the energy--momentum and spin
currents of matter fields, respectively.

The first field equation can be rewritten as
$$
-4\nabla_\m\bigl(b\b_i{^{\m k}}\bigr)+2b\b_{imn}T^{kmn} 
                   -4b\b^{nmk}T_{nmi}+\d_i^k b\cL_T=\t^k{_i}\, .
$$
Then, combined with the identity \eq{A5}, it takes the form of
Einstein's field equation:  
\bsubeq \lab{2.4} 
\be
R^{ik}(\D)-\fr{1}{2}\eta^{ik}R(\D)=\t^{ki}/2ab\, .        \lab{2.4a}
\ee
Here, on the left hand side we have Einstein's tensor of GR, which is a
symmetric tensor. Therefore, the dynamical energy--momentum tensor must
be also symmetric, $\t^{ik}=\t^{ki}$.  

Using the identity \eq{A1} the second field equation can be written in
the form 
\be
\nabla_\m \bigl( 2aH_{ij}^{\n\m}+4b\l_{ij}{^{\n\m}}\bigr)
                               =\s^\n{}_{ij} \, ,         \lab{2.4b}
\ee
\esubeq 
where $H_{ij}^{\n\m}=b(h_i{^\n}h_j{^\m}-h_j{^\n}h_i{^\m})$.
The integrability condition for this equation is identically
satisfied, because the covariant divergence of the left hand side
vanishes on account of $R^{ij}{}_{\m\n}=0$, whereas 
$$
\nabla_\n\s^\n{}_{ij}= \t_{ij}-\t_{ji} =0\, .
$$
The first equality in this relation is a consequence of the covariant
conservation of angular momentum for matter fields (which holds when
matter field equation is satisfied), and the vanishing of $\t_{[ij]}$
follows from the first field equation.  

Simple counting shows that the number of independent field equations
\eq{2.4b} is $24-6=18$. The multipliers $\l_{ij}{}^{\a\b}$ remain
arbitrary functions of time, as will be shown in the forthcoming 
Hamiltonian analysis. For any specific choice of $\l_{ij}{}^{\a\b}$
(gauge fixing), equations \eq{2.4b} can be used to determine (at least
locally) the remaining 18 multipliers $\l_{ij}{}^{0\a}$.  

In what follows we shall investigate the Hamiltonian structure and
gauge properties of \tgr\ without matter fields ($\s^\n{}_{ij}=\t_{ij}=0$). 
We expect that the results obtained here will be also useful for the
analysis of interacting \tgr. 

\section{Primary constraints and Hamiltonian} 

\subsub{1.} The basic Lagrangian dynamical variables of our theory are 
$(b^i{}_\m,A^{ij}{}_\m,\l_{ij}{}^{\m\n})$, and the corresponding
momenta are denoted by $(\p_i{}^\m,\p_{ij}{}^\m,\p^{ij}{}_{\m\n})$.
Due to the fact that the torsion and the curvature do not involve
velocities $\dot b^k{}_0$ and $\dot A^{ij}{}_0$, one immediately obtains
the following set of the so--called {\it sure\/} primary constraints:
\be
\phi_k{}^0\equiv \p_k{}^0\approx 0\, ,\qquad
\phi_{ij}{}^0\equiv \p_{ij}{}^0\approx 0\, .                 \lab{3.1}
\ee

Similarly, the absence of the time derivative of $\l_{ij}{}^{\m\n}$
implies   
\be
\phi^{ij}{}_{\m\n}\equiv \p^{ij}{}_{\m\n} \approx 0\, .      \lab{3.2}
\ee

The next set of constraints follows from the linearity of the curvature
in $\dot A^{ij}{}_\a$:
\be
\phi_{ij}{}^\a\equiv\p_{ij}{}^\a-4b\l_{ij}{}^{0\a}\approx 0\, .\lab{3.3}
\ee

Before we continue, it is convenient introduce the so--called (3+1)
decomposition of spacetime \cite{22}. If \mb{n} is the unit normal to the 
hypersurface $\S_0:$ $x^0=$ const., with $n_k=h_k{^0}/\sqrt{g^{00}}$,
the four vectors $(\mb{n},\mb{e}_\a)$ define the ADM basis of tangent
vector fields. Introducing the projectors on \mb{n} and $\S_0$,
$(P_\ort)^i_k=n^in_k$, $(P_\parallel)^i_k=\d^i_k-n^in_k$, any tangent
vector $\mb{V}$ can be decomposed in terms of its parallel 
and orthogonal components: $V_k=V_\bk+n_kV_\ort$, where
$V_\ort=b^kV_k$, $V_\bk=V_k-n_kV_\ort$, and $n^kV_\bk=0$. Using an
analogous decomposition of the torsion and the curvature in last two
indices,  
$$
T^i{}_{mk}=T^i{}_{\bm\bk}+2n_{[m}T^i{}_{\ort\bk]}\, ,\qquad
R^{ij}{}_{mk}=R^{ij}{}_{\bm\bk}+2n_{[m}R^{ij}{}_{\ort\bk]}\, ,
$$
we find that the parallel components $T^i{}_{\bk\bl}$ and
$R^{ij}{}_{\bk\bl}$ are independent of velocities. The replacement in
the gravitational Lagrangian yields  
$\cL=\bar\cL(T^i{}_{\bk\bl},R^{ij}{}_{\bk\bl};
           T^i{}_{\ort\bl},R^{ij}{}_{\ort\bl},n^k)$.

The decomposition of $\mb{e}_0$ in the ADM basis yields
$\mb{e}_0=N\mb{n}+N^\a\mb{e}_\a$, where $N=n_kb^k{_0}$ and
$N^\a=h_\bk{^\a}b^k{_0}$ are lapse and shift functions, respectively.
We note also that $b$ satisfies the factorization property $b=NJ$, where
$J$ does not depend on $b^k{_0}$.  

Now, we turn our attention to the remaining momenta $\p_i{}^\a$
\cite{22}. The relations defining $\pi_i{^\a}$ can be written in the
form  
$$
\hp_i{^\bk}=J{\pd\bar\cL_T\over\pd T^i{}_{\ort\bk}}
           = 4J\b_i{}^{\ort\bk}(T) \, ,
$$
where $\hp_i{^\bk}=\pi_i{^\a}b^k{_\a}$ are conveniently defined
``parallel" gravitational momenta.  Using now the fact that $\b$ is a
linear function of $T$ we can make the expansion $\b(T)=\b(0)+\b(1)$,
where $\b(0)$ does not depend on ``velocities" $T^i{}_{\ort\bk}$ and
$\b(1)$ is linear in them, and rewrite the above equation in the form
$$
P_{i\bk}\equiv \hp_{i\bk}/J-4\b_{i\ort\bk}(0)=4\b_{i\ort\bk}(1)\, .
$$
Here, the so--called ``generalized momenta" $P_{i\bk}$ do not depend on
velocities, which appear only on the right hand side of the equation.
Explicit calculation leads to the result  
\bea
P_{i\bk}\,&&\equiv \hp_{i\bk}/J 
        -4a\bigl[\fr{1}{2}BT_{\ort\bi\bk}
                +\fr{1}{2}Cn_iT^\bm{}_{\bm\bk}\bigr] \nn \\
        &&=4a\bigl[ AT_{i\ort\bk}+\fr{1}{2}BT_{\bk\ort\bi}  
                  +\fr{1}{2}C\eta_{\bi\bk}T^\bm{}_{\ort\bm}
                  +\fr{1}{2}(B+C)n_iT_{\ort\ort\bk}\bigr]\, . \nn
\eea

This system of equations can be decomposed into irreducible parts with
respect to the group of three--dimensional rotations in $\S_0$:   
\bea
P_{\ort\bk}\,&&\equiv\hp_{\ort\bk}/J-2aCT^\bm{}_{\bm\bk}
              =2a(2A+B+C)T_{\ort\ort\bk} \, , \nn \\
P^A_{\bi\bk}\,&&\equiv \hp^A_{\bm\bk}/J-2aBT_{\ort\bi\bk}
              =2a(2A-B)T^A_{\bi\ort\bk} \, ,  \nn \\
P^T_{\bi\bk}\,&&\equiv \hp^T_{\bi\bk}/J
              =2a(2A+B)T^T_{\bi\ort\bk} \, ,  \nn \\
P^\bm{}_\bm \,&& \equiv \hp^\bm{}_\bm/J 
              =2a(2A+B+3C)T^\bm{}_{\ort\bm} \, ,  \nn
\eea
where $X^A_{\bi\bk}=X_{[\bi\bk]}$,
$X^T_{\bi\bk}=X_{(\bi\bk)}-\eta_{\bi\bk}X^\bn{_\bn}/3$. Taking now into
account the special choice of parameters adopted in equation \eq{2.2},
we recognize here two sets of relations: the first set represents 
{\it extra\/} primary constraints,   
\bsubeq  \lab{3.4} 
\bea
P_{\ort\bk}\,&&=\hp_{\ort\bk}/J +2aT^\bm{}_{\bm\bk}\approx 0\, ,\nn \\ 
P^A_{\bi\bk}\,&&=\hp^A_{\bi\bk}/J-aT_{\ort\bi\bk}\approx 0\, ,\lab{3.4a}
\eea
usually called {\it if--constraints\/}, while the second set gives
nonsingular equations,
\bea
P^T_{\bi\bk}\,&&\equiv \hp^T_{\bi\bk}/J =2aT^T_{\bi\ort\bk} \, ,\nn \\
P^\bm{}_\bm\,&&\equiv \hp^\bm{}_\bm/J =-4aT^\bm{}_{\ort\bm}\, ,\lab{3.4b}
\eea
\esubeq 
which can be solved for velocities.

Further calculations are greatly simplified by observing that both
sets of extra constraints \eq{3.4a} can be represented in a unified
manner as  
\be
\phi_{ik}=\pi_{i\bk}-\pi_{k\bi}+a\nabla_\a B^{0\a}_{ik}\, ,\qquad 
   B^{0\a}_{ik}\equiv\ve^{0\a\b\g}_{ikmn}b^m_\b b^n_\g \, .\lab{3.5}
\ee
This is seen from the fact that relations \eq{3.4a} can be equivalently
written as 
$$
\pi_{i\bk}-\pi_{k\bi}\approx 2aJ\bigl(T_{\ort\bi\bk}-n_iT^\bm{}_{\bm\bk}
     +n_k T^\bm{}_{\bm\bi}\bigr)=2a\nabla_\a H^{0\a}_{ik}\, , 
$$
where the last equality follows from \eq{A1}, and the identity
$2H^{\m\n}_{ik}=-B^{\m\n}_{ik}$. 

\subsub{2.} Having found all the primary constraints, we now proceed to
find the {\it canonical\/} Hamiltonian density \cite{22}:
$$
\cH_c=\pi_i{^\a}\dot b^i{_\a}
      +\fr{1}{2}\pi_{ij}{^\a}\dot A^{ij}{_\a}-b\cL \, .
$$ 
The velocities $\dot b^i{_\a}$ and $\dot A^{ij}{_\a}$ can be
calculated from the relations defining $T^i{}_{0\a}$ and
$R^{ij}{}_{0\a}$:  
\bea
T^i{}_{0\a}\,&&\equiv \pd_0 b^i{_\a}+A^i{}_{m 0}b^m{_\a}-\pd_\a b^i{_\a} 
  -A^i{}_{m\a}b^m{_0}= NT^i{}_{\ort\a}+N^\b T^i{}_{\b\a}\, ,\nn \\
R^{ij}{}_{0\a}\,&&\equiv \pd_0 A^{ij}{_\a} + A^i{}_{m 0}A^{mj}{_\a}
  -\pd_\a A^{ij}{_\a}- A^i{}_{m\a}A^{mj}{_0}
  = NR^{ij}{}_{\ort\a}+N^\b R^{ij}{}_{\b\a} \, . \nn
\eea
After a simple algebra we find that the canonical Hamiltonian can be
written as a linear function of unphysical variables
$(b^k{_0},A^{ij}{_0})$, up to a 3--divergence, 
\bsubeq \lab{3.6} 
\be
\cH_c=N\cH_\ort + N^\a\cH_\a -\fr{1}{2}A^{ij}{_0}\cH_{ij}+\pd_\a D^\a\, ,
                                                             \lab{3.6a}
\ee
where
\bea
\cH_{ij}\,&&= 2\p_{[i}{}^\b b_{j]\b} +\nabla_\a\p_{ij}{}^\a \, ,\nn\\
\cH_\a\,&& = \p_k{}^\b T^k{}_{\a\b}- b^k{}_\a\nabla_\b\p_k{}^\b
             +\fr{1}{2}\p_{ij}{}^\b R^{ij}{}_{\a\b}  \, ,\nn\\
\cH_\ort\,&& =\bigl( \hp_i{^\bk}T^i{}_{\ort\bk}-J\bcL_T 
 -n^k\nabla_\b\p_k{^\b}\bigr)-J\l_{ij}{}^{\bm\bn}R^{ij}{}_{\bm\bn}\, ,\nn\\
D^\a\,&&=b^k{}_0\p_k{}^\a +\fr{1}{2}A^{ij}{}_0\p_{ij}{}^\a\, .\lab{3.6b}
\eea
\esubeq 
Here, $\cH_{ij}$ and $\cH_\a$ are purely kinematical terms whose form
does not depend on the choice of the Lagrangian, and $\cH_\ort$ is the
only dynamical part.

The explicit form of $\cH_\ort$ can be obtained by eliminating
``velocities" $T_{i\ort\bk}$ with the help of the relations defining
momenta $\pi_{i\bk}$. To do that we first rewrite the first two terms
of $\cH_\ort$ in the form   
$$
\hp^{i\bk}T_{i\ort\bk}-J\bcL_T 
           =\fr{1}{2}JP^{i\bk}T_{i\ort\bk}-J\bcL_T(\bT)\, ,
$$
where $\bT_{ikl}=T_{i\bk\bl}$. Then, taking into account the
constraints \eq{3.4a} one finds that $T_{\ort\ort\bk}$ and
$T^A_{\bi\ort\bk}$ are absent from $\cH_\ort$, whereupon the  
relations \eq{3.4b} can be used to eliminate the remaining ``velocities" 
$T^T_{\bi\ort\bk}$ and $T^\bm{}_{\ort\bm}$, leading directly to 
\bsubeq \lab{3.7} 
\be
\cH_\ort =\bigl(\fr{1}{2}P_T^2-J\bcL_T(\bT)-n^k\nabla_\b\p_k{^\b}\bigr)
           -J\l_{ij}{}^{\bm\bn}R^{ij}{}_{\bm\bn} \, ,      \lab{3.7a}
\ee
where
\bea
P_T^2\,&&= {1\over 2aJ}\left( \pi_{(\bi\bk)}\pi^{(\bi\bk)} 
           -{1\over 2}\pi^\bm{_\bm}\pi^\bn{_\bn} \right) \, ,\nn\\
\bcL_T(\bT)\,&&=a\left(\fr{1}{4}T_{m\bn\bk}T^{m\bn\bk} 
       + \fr{1}{2}T_{\bm\bn\bk}T^{\bn\bm\bk}  
       - T^\bm{}_{\bm\bk}T_\bn{}^{\bn\bk} \right) \, .     \lab{3.7b}
\eea
\esubeq 

The general Hamiltonian dynamics of the system is described by the 
{\it total\/} Hamiltonian, which is given as  
\bea
\cH_T=\cH_c && +u^i{_0}\p_i{^0}+\fr{1}{2}u^{ij}{_0}\p_{ij}{^0} 
              +\fr{1}{4}u_{ij}{}^{\m\n}\p^{ij}{}_{\m\n}  \nn\\
  &&+\fr{1}{2}u^{ik}\phi_{ik}
              +\fr{1}{2}u^{ij}{_\a}\phi_{ij}{^\a}\, ,     \lab{3.8}
\eea
where $u'$s are, at this stage, arbitrary Hamiltonian multipliers.

Although the torsion components $T_{\ort\ort\bk}$ and
$T^A_{\bi\ort\bk}$ are absent from the canonical Hamiltonian,
they re--appear in the total Hamiltonian as the non--dynamical
Hamiltonian multipliers. Indeed, the Hamiltonian field equations for 
$b^k{_\a}$ imply (Appendix B) 
\be
NT^{\ort\ort\bk}=u^{\ort\bk}\, ,\qquad NT_A^{\bi\ort\bk}=u^{\bi\bk}\, .
                                                          \lab{3.9}
\ee    
The presence of non--dynamical torsion components does not imply that 
\tgr\ is an inconsistent theory \cite{9}, as it has very clear
interpretation via the gauge structure of the theory: it is related to
the existence of additional first--class constraints $\phi_{ik}$, as we
shall see in Sec. V. 

\section{Consistency conditions} 

The consistency of the theory requires that the constraints do not
change during the time evolution of the system governed by the total
Hamiltonian: 
$$
\chi(x)\equiv{d\over dt}\phi(x)=\int d^3x'\{\phi,\cH'_T\}\approx 0\, ,
$$
where $\{A,B'\}$ denotes the Poisson bracket (PB) of two variables
$A(x)$ and $B(x')$, and $x^0=(x')^0$.  The integration sign will be
often omitted for simplicity.

\subsection{Consistency conditions of primary constraints} 

Having found the form of primary constraints in \tgr, displayed in Eqs.
\eq{3.1}, \eq{3.2}, \eq{3.3} and \eq{3.5}, we now consider the
requirements for their consistency.  

Since the canonical Hamiltonian is linear in unphysical variables
$(b^i{_0},A^{ij}{_0})$, the consistency conditions of the sure primary
constraints \eq{3.1} are given by  
\be
\chi_\ort\equiv \cH_\ort\approx 0\, ,\qquad
\chi_\a  \equiv \cH_\a\approx 0\, ,\qquad
\chi_{ij}\equiv \cH_{ij}\approx 0\, .                       \lab{4.1}
\ee

By noting that the components $\p^{ij}{}_{\a\b}$ in Eq.\eq{3.2} have 
vanishing PBs with all primary constraints, we easily obtain   
\bsubeq \lab{4.2} 
\be
\chi^{ij}{}_{\a\b}\equiv R^{ij}{}_{\a\b}\approx 0 \, .      \lab{4.2a}
\ee
On the other hand, the consistency of $\p^{ij}{}_{0\a}$ implies 
\be
\chi^{ij}{}_{0\b}\equiv u^{ij}{_\b}-N^\a R^{ij}{}_{\a\b}
   \approx 0 \quad \Ra \quad u^{ij}{_\b}\approx 0 \, .      \lab{4.2b}
\ee
\esubeq 
The dynamical meaning of the last condition can be seen more clearly
if we note that the equation of motion for $A^{ij}{_\b}$, 
$\pd_0 A^{ij}{_\b}=\{ A^{ij}{_\b}, H_c\} +u^{ij}{_\b}$, 
can be transformed into the form
\be
R^{ij}{}_{0\b}\approx u^{ij}{_\b} \, .                       \lab{4.3}
\ee
Hence, Eqs. \eq{4.2} tell us that all the components of the curvature
tensor weakly vanish, as one could have expected.

Using the PB relation
$$
\{\phi_{ij}{^\a},\phi_{kl}\} =a\bigl( \eta_{ik}B^{0\a}_{lj}
                    +\eta_{jk}B^{0\a}_{il}\bigr)\d -(kl)\, ,
$$
the consistency condition for $\phi_{ij}{}^\a$ takes the form 
$$
\chi_{ij}{^\a}= \{\phi_{ij}{^\a},\cH_c\} 
  + a\bigl( u_i{^s}B^{0\a}_{sj} +u_j{^s}B^{0\a}_{is} \bigr) 
                   - 4bu_{ij}{}^{0\a} \approx 0 \, .     
$$
It can be used to determine $u_{ij}{}^{0\a}$:
\be
u_{ij}{}^{0\a}={a\over 4b}\bigl( u_i{^s}B^{0\a}_{sj}
   +u_j{^s}B^{0\a}_{is} \bigr) + \bar u_{ij}{}^{0\a}\, ,\qquad
\bar u_{ij}{}^{0\a}\equiv {1\over 4b}\{\pi_{ij}{^\a},\cH_c\}\, ,
                                                          \lab{4.4}
\ee
where $\{\pi_{ij}{^\a},\cH_c\}$ is calculated in Appendix C.
The first part of $u_{ij}{}^{0\a}$ contains $u_{kl}$ and gives an
additional contribution to $u^{kl}\phi_{kl}$, so that the replacement
of this result into $\cH_T$ leads effectively to 
\bea
&& u_{ij}{}^{0\a} \to \bar u_{ij}{}^{0\a}\, ,\qquad
   u^{kl}\phi_{kl}\to u^{kl}\tphi_{kl}\, ,\nn\\
&& \tphi_{kl}\equiv\phi_{kl}-{a\over 4b} \bigl( 
   \pi_k{^s}{}_{0\a}B_{sl}^{0\a}+\pi_l{^s}{}_{0\a}B_{ks}^{0\a}\bigr)\, .
                                                           \lab{4.5}
\eea
Note that $\{\phi_{ij}{^\a},\tphi'_{kl}\}=0$.

The most complicated consistency conditions are those for the tetrad
constraints $\phi_{ij}$ (or, equivalently, $\tphi_{ij}$). 
First we note that their PB algebra has the form
\be
\{\phi_{ij},\phi'_{mn}\}= \bigl( \eta_{im}\phi_{nj}
              +\eta_{jm}\phi_{in} \bigr)\d -(mn) \, ,      \lab{4.6}
\ee
and that the term $\fr{1}{2}u^{ij}{_\a}\phi_{ij}{^\a}$ in $\cH_T$ can
be discarded according to \eq{4.2b}. Then, after showing that the
Poisson bracket $\{\phi_{ij},\cH'_c\}$ vanishes weakly, we will be able
to conclude that the consistency condition for $\phi_{ij}$ is
automatically fulfilled:    
\be
\chi_{ij}\equiv \{\phi_{ij}, \cH'_T \}
         \approx \{\phi_{ij}, \cH'_c \} \approx 0 \, .      \lab{4.7}
\ee

\subsection{The consistency condition of \mb{\phi_{ij}}} 

In order to simplify the derivation of the consistency condition for
$\phi_{ij}$ we rewrite this constraint in the form
\bea
\phi_{ij}\,&&=\cH_{ij}-F_{ij} \, ,  \nn\\
F_{ij}\,&&=\nabla_\a\bigl( \pi_{ij}{^\a}-aB_{ij}^{0\a} \bigr)
       \equiv \nabla_\a\Pi_{ij}{^\a}\, .                  \lab{4.8}
\eea
General arguments in PGT, related to the local Lorentz symmetry of the
theory, imply that the constraint $\cH_{ij}$ is of the first class
\cite{22}. This is also clear from the PB algebra of constraints,
discussed in the next section. As a consequence, the consistency of
$\phi_{ij}$ follows from the consistency of $F_{ij}$. 

We are now going to show that $\{F_{ij},\cH'_T\}\approx 0$. First, we
note that  
\be
\{ F_{ij},\phi'_{kl}\}=0 \, .                              \lab{4.9}
\ee
Then, using the results $(C.4a)$ we obtain
\bsubeq  \lab{4.10} 
\bea
&&\{ F_{ij},\cH'_{kl} \}=
      \left(\eta_{ik}F_{lj} + \eta_{jk} F_{il}\right)\d -(kl)\, ,\nn\\
&&\{ F_{ij}, \cH'_\b \} = -\nabla_\b\bigl( \phi_{ij}\d\bigr)
 +[\nabla_\a,\nabla_\b]\bigl(\Pi_{ij}{^\a}\d\bigr) \, .   \lab{4.10a}
\eea
while $(C.4b)$ leads to
\bea
&&\{F_{ij},\cH'_\ort\}= \fr{1}{2}\nabla_\b\bigl[ \bigl(
              n_i\phi_{jk}-n_j\phi_{ik}\bigr)h^{\bk\b}\d \bigr]
  + \fr{1}{2} [\nabla_\a,\nabla_\b]\bigl(M_{ij}^{\a\b}\d\bigr)\, ,\nn\\
&& NM_{ij}^{\a\b}\equiv  2aH^{\a\b}_{ij} -4b\l_{ij}{}^{\b\a} 
  +N^\a\bigl( \Pi_{ij}{}^{\b}-\phi_{ij}{^\b}\bigr)
  -N^\b\bigl( \Pi_{ij}{}^{\a}-\phi_{ij}{^\a}\bigr)  \, ,   \lab{4.10b}
\eea
\esubeq
Hence, the consistency condition of $F_{ij}$, and consequently of
$\phi_{ij}$, is automatically satisfied.

\subsection{Consistency conditions of secondary constraints} 

In the process of investigating the consistency of primary constraints
in \tgr, we obtained secondary constraints \eq{4.1} and \eq{4.2a}.

Consider, first, the consistency condition of the secondary constraint
$R^{ij}{}_{\a\b}$. Since $R^{ij}{}_{\a\b}$ depends only on
$A^{ij}{}_\a$, one can express $dR^{ij}{}_{\a\b}/dt$ in terms of
$dA^{ij}{_\a}/dt$, use the equation of motion \eq{4.3} for $A^{ij}{_\a}$,
and rewrite the result in the form 
$$
\nabla_0 R^{ij}{}_{\a\b}\approx 
       \nabla_\a u^{ij}{_\b}-\nabla_\b u^{ij}{_\a} \, .    
$$
Hence, the consistency condition for $R^{ij}{}_{\a\b}$ is identically
satisfied.    

The above relation has a very interesting geometric interpretation.
Indeed, using  equation \eq{4.3} we see that it is a weak consequence of
the second Bianchi identity.    

General arguments in PGT show that the secondary constraints
$\cH_{ij},\cH_\a$ and $\cH_\ort$ are related to Poincar\'e gauge
symmetry \cite{22,24}. Consequently, they are of the first class, and
their consistency conditions are automatically satisfied. This will be
explicitly seen in the next section, from the form of their PB algebra. 

Finally, at the end of the consistency procedure, we give the final
expression for the total Hamiltonian: 
\bea
\cH_T\,&&=\cH'  +u^i{_0}\p_i{^0}+\fr{1}{2}u^{ij}{_0}\p_{ij}{^0} 
             +\fr{1}{4}u_{ij}{}^{\a\b}\p^{ij}{}_{\a\b} 
             +\fr{1}{2}u^{ik}\tphi_{ik}        \, ,\nn\\
\cH'\,&&=\cH_c+\fr{1}{2}{\bar u}_{ij}{}^{0\b}\p^{ij}{}_{0\b}\, .\lab{4.11}
\eea
The multipliers $u^i{_0},u^{ij}{_0},u_{ij}{}^{\a\b}$ and $u^{ik}$
remained arbitrary functions of time, hence we expect that the related
constraints $\p_i{^0},\p_{ij}{^0},\p^{ij}{}_{\a\b}$ and $\tphi_{ik}$
are of the first class. 

\section{The algebra of constraints} 

In the previous analysis we found that \tgr\ is characterized by the
following set of constraints:  
\bitem
\item[] primary:$\quad\pi_i{^0},\pi_{ij}{^0},\tphi_{ij},
           \pi^{ij}{}_{\a\b},\phi_{ij}{^\a},\pi^{ij}{}_{0\b}\, ;$ \\
 secondary:$\quad\cH_\ort,\cH_\a,\cH_{ij}, R^{ij}{}_{\a\b}\, .$
\eitem
It is simple to see that $\phi_{ij}{^\a}$ and $\pi^{ij}{}_{0\b}$
are {\it second class\/} constraints. They can and will be used as
strong equalities to eliminate $\l_{ij}{}^{0\a}$ and $\pi^{ij}{}_{0\b}$
from the theory and simplify further exposition. In particular, the
term with the determined multiplier $\bar u^{ij}{_\a}$ in the total
Hamiltonian can be now neglected, and $\tphi_{ij}$ reduces to
$\phi_{ij}$, equation \eq{4.5}. Due to a simple form of these second
class constraints, the related Dirac brackets have the form of PBs in
the phase space  of the remaining variables.  All the remaining
constraints are of the {\it first class\/,} as follows from their PB
algebra.

Since the kinematical constraints $\cH_{ij},\cH_\a,\cH_\ort$ have the 
same general form as in \cite{22,24}, their algebra remains the same:
\bsubeq \lab{5.1} 
\bea
 &&\{\cH_{ij},\cH'_{kl}\} =\bigl( \eta_{ik}\cH_{lj}
                       +\eta_{jk}\cH_{il} \bigr)\d -(kl)\, ,\nn\\
 &&\{\cH_{ij},\cH'_\a \} = 0 \, ,\nn\\
 &&\{\cH_\a ,\cH'_\b \} =\bigl(\cH_\a '\pd_\b +\cH_\b\pd_\a 
       -\fr{1}{2}R^{ij}{_{\a\b}}\cH_{ij}\bigr)\d\, .        \lab{5.1a}
\eea
As a consequence of $R^{ij}{}_{\a\b}\approx 0$, the
last term in $\{\cH_\a ,\cH'_\b \}$ is quadratic in constraints. 

The brackets involving $\cH_\ort$ are found to have the form 
\bea
 &&\{\cH_{ij},\cH'_\bot \} =0 \, ,\nn\\
 &&\{\cH_\a ,\cH'_\bot \} =\cH_\bot\pd_\a \d \, ,\nn\\
 &&\{\cH_\bot ,\cH'_\bot \} =-\bigl({^3}g^{\a\b}\cH_\a +
         {^3}g^{\prime\a\b}\cH_\a '\bigr)\pd_\b\d\, .       \lab{5.1b}
\eea
\esubeq 

The first two brackets are most easily verified by taking into account
that $\cH_\ort$ can be written in the form 
$\cH_\ort=Jf(\xi^A)-n^k\nabla_\a\pi_k{^\a}$, where $f$ is a Lorentz
scalar formed from variables $\xi^A=(T^i{}_{\bm\bn},
R^{ij}{}_{\bm,\bn},\hp_i{^\bk}/J,\hp_{ij}{^\bk}/J, n^k)$, as shown in
\cite{24}. The second bracket is obtained from the general formula
based on the chain rule for PBs, 
$$
\{\cH_\a,\cH'_\ort\}={\pd\cH_\a\over\pd\xi^A}\{\xi^A,\xi'^B\}
        {\pd\cH'_\ort\over\pd\xi'^B}= \left(\cH_\bot\pd_\a
  +\fr{1}{2}{\pd\cH_\ort\over\pd\pi_{ij}{^\a}}\cH_{ij}\right)\d \, ,
$$
which explains why the second term is absent in \eq{5.1b}.

The last and most important bracket $\{\cH_\ort,\cH'_\ort\}$  is
evaluated using the chain rule and keeping only those terms that
contain $\pd_\a\d$ (terms proportional to $\d$ do not have the 
correct symmetry under $\mb{x}\lra\mb{x}'$, hence they cancel each
other).    

In the next step we want to extend the above algebra by adding
$\tphi_{ij}=\phi_{ij}$. The relevant PBs involving $\phi_{ij}$ are
given by   
\bea
&&\{\phi_{ij},\phi_{kl}\}=(\eta_{ik}\phi_{lj}+\eta_{jk}\phi_{il}) 
                            -(kl)\, ,\nn\\
&&\{\phi_{ij},\cH'_{kl}\} =\bigl( \eta_{ik}\phi_{lj}
                       +\eta_{jk}\phi_{il} \bigr)\d -(kl)\, ,\nn\\
&&\{\phi_{ij}, \cH'_\b \} = \nabla_\b\bigl( \phi_{ij}\d\bigr)
 -[\nabla_\a,\nabla_\b]\bigl(\Pi_{ij}{^\a}\d\bigr) \, ,\nn\\
&&\{\phi_{ij},\cH'_\ort\}= -\fr{1}{2}\nabla_\b\bigl[ \bigl(
          n_i\phi_{jk}-n_j\phi_{ik}\bigr)h^{\bk\b}\d \bigr]
  -\fr{1}{2}[\nabla_\a,\nabla_\b]\bigl(M_{ij}^{\a\b}\d\bigr)\, .
                                                        \lab{5.2}
\eea

Finally, we display the non--vanishing PBs involving $R^{ij}{}_{\a\b}$
and $\pi^{ij}{}_{\a\b}$: 
\bea
&&\{R^{ij}{}_{\a\b},\cH'_{kl}\}=(\d^i_k R_l{^j}{}_{\a\b}+
                              \d^j_k R^i{_l}{}_{\a\b})\d-(kl)\, ,\nn\\
&&\{R^{ij}{}_{\a\b},\cH'_\g\}=\nabla_\a(R^{ij}{}_{\g\b}\d)-(\a\b)\, ,\nn\\
&&\{\pi^{ij}{}_{\a\b},\cH'_\ort\}=4JR^{ij}{}_{\a\b}\, .      \lab{5.3}
\eea

Thus, all constraints except $\phi_{ij}{^\a}$ and $\pi^{ij}{}_{0\b}$
are of the first class. The fact that $\phi_{ij}$ is first class
is of particular importance for the consistent interpretation of the
non--dynamical torsion components, as noted at the end of Sec. III.

\section{Extra gauge symmetries} 

The presence of arbitrary multipliers in the total Hamiltonian is
related to the existence of gauge symmetries in the theory.
The general method of constructing the generators of such symmetries
has been given by Castellani \cite{25}. If we limit ourselves to gauge
transformations given in terms of arbitrary parameters $\ve(t)$ and
their first time derivative $\dot\ve(t)$, which is sufficient for the
present analysis, the gauge generators take the form 
\bsubeq \lab{6.1} 
\be
G=\int d^3x\bigl[ \ve(t)G^\nul +\dot\ve(t)G^\one \bigr]\, ,  \lab{6.1a}
\ee 
where $G^\nul$ and $G^\one$ are phase space functions determined by the
conditions
\bea
G^\one\,&& =C_{PFC}\, ,\nn\\
G^\nul+\{ G^\one,H_T\}\,&&=C_{PFC}\, ,\nn\\
\{ G^\nul,H_T\}\,&&=C_{PFC} \, ,                            \lab{6.1b}
\eea
\esubeq
and $C_{PFC}$ denotes primary first class (PFC) constraint.

The Poincar\'e  gauge symmetry is present in our formulation of \tgr\
by construction, and the related gauge generator is based on the sure
constraints $\pi_i{^0},\pi_{ij}{^0}$ and $\cH_\ort,\cH_\a,\cH_{ij}$  
\cite{26}. Here, we shall focus our attention on extra gauge
symmetries based on $\pi^{ij}{}_{\a\b},\tphi_{ij}$ and $R^{ij}{}_{\a\b}$.

\subsection{Extra gauge symmetry, 1} 

Starting with $\pi^{ij}{}_{\a\b}$ as $G^\one$ in \eq{6.1b} we find that
the related gauge generator is given by
\be
G=\int d^3x\left[\fr{1}{4}\bigl( 
     \nabla_0\ve_{ij}{}^{\a\b}\bigr) \pi^{ij}{}_{\a\b}
    +\fr{1}{4}\ve_{ij}{}^{\a\b}\bigl( -4bR^{ij}{}_{\a\b}
                +(\dot b/b)\pi^{ij}{}_{\a\b}\bigr)\right]\, .\lab{6.2}
\ee
The only nontrivial gauge transformations $\d_0 X=\{ X,G\}$ are 
\bea
&& \d_0(b\l_{ij}{}^{\a\b})= \nabla_0(b\ve_{ij}{}^{\a\b})\, ,\nn\\
&& \d_0 \pi_{ij}{^\a} = 4\nabla_\b (b\ve_{ij}{}^{\a\b})\, .  \lab{6.3}
\eea

To see the meaning of these transformations, consider the Hamiltonian
equation for the variable $\Pi_{ij}{^\a}=\pi_{ij}{^\a}-aB_{ij}^{0\a}$.
Introducing $K_{ij}^{\a\b}=4b\l_{ij}{}^{\a\b}-aB_{ij}^{\a\b}$ and using
the results of Appendix C we obtain the equation
\be
\nabla_0\Pi_{ij}{^\a} -\nabla_\b K_{ij}^{\a\b} =0\, ,       \lab{6.4}
\ee
which is the Hamiltonian analogue of \eq{2.4b}. The application of the
above gauge transformation to this equation yields 
$$
(\nabla_0\nabla_\b-\nabla_\b\nabla_0)(4b\ve_{ij}{}^{\a\b})=0 \, .
$$
The invariance follows from the fact that the left hand side vanishes
in Weitzenb\"ock space, where $R^{ij}{}_{0\b}=0$.

\subsection{Extra gauge symmetry, 2} 

Starting with $G^\one_{ij}=\phi_{ij}$ in \eq{6.1b}, one finds that the
gauge generator has the form  
\bsubeq \lab{6.5} 
\be
G_{ij}=\int d^3x \bigl[ \fr{1}{2}\dot\ve^{ij}G^\one_{ij} 
               +\fr{1}{2}\ve^{ij}G^\nul_{ij} \bigr] \, ,  \lab{6.5a}
\ee 
where (Appendix D)
\be
G^\nul_{ij}={1\over 2} R_i{^s}{}_{\a\b} K_{sj}^{\a\b}+
{1\over 2}{1\over 4b}\left[  
 \left( A_i{^n}{_0} \pi_n{^s}{}_{\a\b} 
   +A^s{}_{n0}\pi_i{^n}{}_{\a\b} \right) K_{sj}^{\a\b} 
   -\pi_i{^s}{}_{\a\b}\dot K_{sj}^{\a\b}\right] -(ij)\, . \lab{6.5b}
\ee
\esubeq 

The corresponding gauge transformations are:
\bea
&& \d_0b^k{_\a}=\dot\ve^k{_s}b^s{_\a} \, ,\qquad
  \d_0A^{ij}{_\a}=0\, ,\nn\\
&& 4b\,\d_0 \l_{ij}{}^{\a\b}= \bigl[ \ve_i{^n}\dot K_{nj}^{\a\b} 
  + A_i{^m}{_0} \bigl( \ve_m{^n} K_{nj}^{\a\b}+
    \ve_j{^n} K_{mn}^{\a\b}\bigr)\bigr] -(ij) \, ,\nn\\
&&\d_0 \pi_{ij}{^\a}= \bigl[ a\bigl( \dot\ve_i{^s}B_{sj}^{0\a}\bigr)
  +\nabla_\b\bigl(\ve_i{^n}K_{nj}^{\a\b}\bigr)\bigr]-(ij)\, ,\lab{6.6}
\eea
and similarly for other variables.

Consider, again, equation \eq{6.4}. Using
$\d_0\Pi_{ij}{^\a}=\nabla_\b\bigl(\ve_i{^n}K_{nj}^{\a\b}\bigr)-(ij)$, 
we easily obtain
\bea
\d_0\bigl(\nabla_0\Pi_{ij}{^\a}\bigr)\,&&
   =\nabla_0\bigl( \d_0\Pi_{ij}{^\a}\bigr)
   \approx\nabla_\b\nabla_0\bigl(\ve_i{^n}K_{nj}^{\a\b}\bigr)-(ij)\nn\\
 &&=\nabla_\b\bigl[ 
  \dot\ve_i{^n}K_{nj}^{\a\b}+\ve_i{^n}\dot K_{nj}^{\a\b}
  +A_i{^s}{_0}\bigl(
   \ve_s{^n}K_{nj}^{\a\b}+\ve_j{^n}K_{sn}^{\a\b}\bigr)\bigr]-(ij)\, ,\nn 
\eea
where we made use of $R^{ij}{}_{0\b}=0$. On the other hand,
$$
\d_0 K_{ij}^{\a\b}= \bigl[ \dot\ve_i{^n}K_{nj}^{\a\b}
    +\ve_i{^n}\dot K_{nj}^{\a\b}  +A_i{^s}{_0}\bigl(
   \ve_s{^n}K_{nj}^{\a\b}+\ve_j{^n}K_{sn}^{\a\b}\bigr)\bigr]-(ij) \, ,
$$
and we see that equation \eq{6.4} is gauge invariant.

\section{Concluding remarks} 

The investigation of the Hamiltonian structure of the teleparallel
formulation of GR presented here is based on Dirac's general method for
constrained dynamical systems \cite{20}.

To complete our results, we now discuss how the physical degrees
of freedom of \tgr\ are counted. After the elimination of
$\l_{ij}{}^{0\a}$ and $\pi^{ij}{}_{0\a}$, the reduced phase space is
spanned by the $40+18$ field components
$(b^i{_\m},A^{ij}{_\m},\l_{ij}{}^{\a\b})$, and the same number of
momenta.  The primary first class constraints $\pi^{ij}{}_{\a\b}$
diminish the number of independent variables for $2\times 18$,
leaving us with the phase space containing effectively $2\times 40$ 
components. 
Before going on, we wish to clarify the counting of constraints
$R^{ij}{}_{\a\b}\approx 0$. Note that here we have formally 18
equations, but they represent only 12 independent conditions on 
$A^{ij}{_\a}$. Indeed, starting with the simplest solution 
$\bar A^{ij}{_\a}=0$ of $R^{ij}{}_{\a\b}(A)=0$, one can construct a
new, Lorentz rotated solution 
$\bar A^{ij}{_\a}(\L)=\L^i{_k}\pd_\a\L^{jk}$, containing 6 arbitrary
parameters $\L^{ik}$ \cite{8}, so that the number of independent conditions
on $A^{ij}{_\a}$ is $18-6=12$. Continuing now the counting, we find  20
sure first class constraints [ten primary $(\pi_i{^0},\pi_{ij}{^0})$,
and ten secondary $(\cH_\ort,\cH_\a,\cH_{ij})$], and $6+12=18$ additional
first class constraints $\phi_{ij}$ and $R^{ij}{}_{\a\b}$, which leaves
us with $2\times 40 -2\times 38=4$ physical degrees of freedom, 
corresponding to massless graviton.

We found two types of extra gauge symmetries in the PGT formulation
of \tgr. The first type is related to the primary constraints 
$\pi^{ij}{}_{\a\b}$. The related gauge transformations do not act on
$b^i{_\m}$, hence they are irrelevant for the structure of the
first field equation \eq{2.4a}. On the other hand, the gauge symmetry
acts nontrivially on Lagrange multipliers. If we recall that the only
role of the second field equation \eq{2.4b} is to determine these
multipliers \cite{9}, it becomes clear that this cannot be done
uniquely without fixing the gauge.  

The second type of extra gauge symmetry originates from the tetrad
constraints $\phi_{ij}$. We note that Nester \cite{14} 
derived these constraints in the form \eq{3.4a}, in his analysis of the
positivity of energy in the teleparallel form of \tgr. Their existence
may be interpreted as a consequence of the fact that the velocities
contained in $T_{\ort\ort\bk}$ and $T^A_{\bi\ort\bk}$ appear at most
linear in the Lagrangian \cite{15} and, consequently, remain
arbitrary functions of time.  The phenomenon that some velocities are
dynamically undetermined is quite usual for constrained dynamical
systems \cite{20}. For the related initial value problem to be well
defined, these undetermined velocities should be {\it removed\/} from
the set of dynamical velocities \cite{15}.     

The role of this symmetry is very clearly seen if we observe that the
teleparallel geometry can be also formulated as the translational gauge
theory, where local Lorentz symmetry is in general absent \cite{5,8}.
However, for the special choice of parameters corresponding to \tgr\
one finds that $\phi_{ij}$ is an additional first class constraint,
which generates local Lorentz symmetry as an extra gauge symmetry
\cite{14}. This also clarifies the form \eq{3.5} of $\phi_{ij}$, which
is seen to ``imitate" $\cH_{ij}$ in the tetrad sector.      

Maluf \cite{21} studied \tgr\ by imposing the time gauge at the
Lagrangian level. His arguments concerning the necessity of the time
gauge in the canonical formalism are conceptually misleading: this
gauge (as well as any other gauge) may be useful, but certainly not
essential \cite{20}. After fixing the time gauge, he found the
Hamiltonian and derived the constraint corresponding to our
$\phi_{\bi\bk}$ [Eq. (25) in his paper], while $\phi_{\ort\bk}$ is
missed. Moreover, Maluf was not able to calculate the constraint
algebra unless imposing another gauge condition. His constraint algebra
[Eqs. (30)-(34)] does not agree with our results, which might be a
consequence of the adopted gauge conditions. All this makes this
analysis of the gauge structure of \tgr\ rather unclear.

The results obtained in this paper refer to non--interacting \tgr, 
and can be used to define and analyze the gravitational energy and
other conserved quantities \cite{27,16}. Interaction with matter fields
may be included in a straightforward manner \cite{7,28}. Studying 
consistency requirements imposed by extra gauge symmetries on the
matter sector will tell us more about the existence and nature of
consistent couplings \cite{15}. 

\section*{Acknowledgment} 

This work was partially supported by the Serbian Science Foundation,
Yugoslavia. 

\appendix 

\section{Some geometric identities in \mb{T_4}} 

We begin with a simple but technically important identity
\bea
\nabla_\n H^{\m\n}_{ij}\,&&=bh_k{^\m}\bigl( T^k{_{ij}} 
    -\d^k_i T_j +\d^k_jT_i \bigr)=-4b\b_{[ij]}{^\m}/a \, ,\nn\\
H^{\m\n}_{ij}\,&&\equiv b\bigl(h_i{^\m}h_j{^\n}-h_j{^\m}h_i{^\n}\bigr)\, ,
                                                             \lab{A1}
\eea
which implies $\nabla_\m(b\b_{[ij]}{^\m})=0$ for $R^{ij}{}_{\m\n}=0$.

In Riemann--Cartan space $U_4$ the Lorentz connection can be expressed 
in the form $A=\D+K$, where $\D$ is Levi--Civita connection and $K$ the
contortion. Substituting this expression into the definition of the
curvature tensor $R^{ij}{}_{\m\n}(A)$, we obtain the basic identity:  
\be
R^{ij}{_{\m\n}}(A)=R^{ij}{_{\m\n}}(\D) +\left[ 
  \nabla'_\m K^{ij}{_\n} +K^i{_{s\m}}K^{sj}{_\n} -(\m\n)\right]\, ,
                                                             \lab{A2}
\ee
where $\nabla'=\nabla(\D)$ is Riemannian covariant derivative. Then,
multiplying this relation with $H_{ij}^{\m\n}/2$ and using 
$\nabla'_\m H_{ij}^{\m\n}=0$, we find 
\be
bR(A)=bR(\D)+b\bigl(\fr{1}{4}T_{ijk}T^{ijk}
   +\fr{1}{2}T_{ijk}T^{jik}-T_kT^k \bigr)+2\pd_\m(bK^\m)\, ,\lab{A3}
\ee
where $K^\m=K^{\m n}{_n}=-T^\m$.

Now, if we write equation $(A2)$ in an equivalent form,
$$
R^{ij}{_{\m\n}}(A)=R^{ij}{_{\m\n}}(\D) +\left[ 
  \nabla_\m K^{ij}{_\n} -K^i{_{s\m}}K^{sj}{_\n} -(\m\lra\n)\right]\, ,
$$
and multiply it with $H_{kj}^{\m\n}/2$, we obtain the result
$$
bR^i{_k}(A)=bR^i{_k}(\D) +\left[
  \nabla_\m K^{ij}{_\n} -K^i{_{s\m}}K^{sj}{_\n}\right]H_{jk}^{\n\m}\, ,
$$
which can be written as
\bea
abR^{ik}(A)=&&abR^{ik}(\D)+ 2\nabla_\m(b\b^{i\m k})
                          + 2b\b_{mn}{^k}T^{mni} \nn\\
          &&-b\b^{imn}T^k{}_{mn} -\eta^{ik}a\pd_\m(bT^\m)
           -4\nabla_\m(b\b^{[ik]\m})\, .                     \lab{A4}
\eea
The last term on the right hand side vanishes for $R^{ij}{}_{\m\n}(A)=0$.
In that case we find
\be
2ab\bigl[ R^{ik}(\D)-\fr{1}{2}\eta^{ik}R(\D)\bigr]= 
          -4\nabla_\m(b\b^{i\m k})-4b\b_{mn}{^k}T^{mni} 
          +2b\b^{imn}T^k{}_{mn} +\eta^{ik}b\cL_T \, .         \lab{A5}
\ee

\section{Unphysical torsion components} 

In this Appendix we show that the unphysical torsion components
$T_{\ort\ort\bk}$ and $T^A_{\bi\ort\bk}$ can be expressed in terms of
the Hamiltonian multipliers $u_{kl}$.  

Using the PB relations
\bea
&& \{b^i{_\a},\cH'_{kl}\}=\d^i_k b_{l\a}\d -(kl)\, , \nn\\
&& \{b^i{_\a},\cH'_\b\}=(\nabla_\a b^i{_\b})\d -b^i{_\b}\pd'_\a\,\d
                       =T^i_{\b\a}\d +\nabla_\a(b^i{_\b}\d) \, , \nn\\
&& \{b^i{_\a},\cH'_\ort\}={1\over 2a}\left( b_{m\a}P_T^{\bi\bm}
         -{1\over 6}b^i{_\a}P^\bm{_\bm} \right)\d
         +\nabla_\a(n^i\d) \, ,                          
\eea
one easily finds that the Hamiltonian equation for $b^k{_\a}$ can be
written in the form
\bea
\nabla_0 b^i{_\a} =&&\nabla_\a b^i{_0} + N^\b T^i{_{\b\a}}\cr
  &&+ Nb_{k\a} {1\over 2aJ}\left( \hp^{(\bi\bk)}
                 -\fr{1}{2}\eta^{\bi\bk}\hp^\bm{_\bm}\right)
    + b_{k\a}\left(n^iu^{\ort\bk}+u^{\bi\bk}\right)\, .
\eea
As a consequence,  
\bea
&&\hp^{(\bi\bk)}-\fr{1}{2}\eta^{\bi\bk}\hp^\bm{_\bm}
                     = 2aJT^{(\bi\ort\bk)}\, ,\nn\\
&&u^{\ort\bk}=NT^{\ort\ort\bk}\, ,\qquad 
                     u^{\bi\bk}=NT_A^{\bi\ort\bk}\, .
\eea

\section{Consistency conditions} 

We collect here several technical relations which simplify the
derivation of the consistency conditions for $\phi_{ij}{^\a}$
and $F_{ij}$.   

\subsub{1.} The term $\{\pi_{ij}{^\a},H_c\}$ in the consistency
condition for the primary constraint $\phi_{ij}{^\a}$ is calculated
using the relations
\bsubeq 
\bea
&&\{ \pi_{ij}{^\a},\cH'_{kl} \}=
     (\eta_{ik}\pi_{lj}{^\a}+ \eta_{jk}\pi_{il}{^\a})\d -(kl)\, ,\nn\\
&&\{ \pi_{ij}{^\a},\cH'_\b \} = -\d^\a_\b(\pi_{i\bj}-\pi_{j\bi})\d 
 -\d^\a_\b\nabla_\g(\pi_{ij}{^\g}\d)+\nabla_\b(\pi_{ij}{^\a}\d)\, . 
                                                          \lab{C1a}
\eea
and
\bea
\{ \pi_{ij}{^\a},\cH'_\ort\}=&&
        -4\nabla_\b(J\l_{ij}{}^{\b\a}\d) 
        +4\nabla_\b\bigl[J(N^\a\l_{ij}{}^{0\b}
        -N^\b\l_{ij}{}^{0\a})\d\bigr]\, ,\nn   \\ 
   && -\left[ 8J\b_{[\bi\bj]\bk}(0)
        + 2aJ n_{[i}T_{\ort\bj]\bk}\right]h^{\bk\a}\d\, ,\nn\\
   && +(n_i\pi_{j\bk}-n_j\pi_{i\bk})h^{\bk\a}\d\, .        \lab{C1b}
\eea
\esubeq 
Using the identity
$8J\b_{[\bi\bj]\bk}(0)h^{\bk\a}=a\ve^{0\a\b\g}_{ijmn}T^m{}_{\g\b} n^n$,
we obtain
\bea
\{\pi_{ij}{^\a},H_c\}=&& 
        -(A_i{^s}{_0}\pi_{sj}{^\a} + A_j{^s}{_0}\pi_{is}{^\a})
        -N^\a(\pi_{i\bj}-\pi_{j\bi})      \nn\\   
 &&-4\nabla_\b(b\l_{ij}{}^{\b\a})-aN\ve^{0\a\b\g}_{ijmn}T^m{}_{\g\b}n^n 
  +N(n_i\pi_{(\bj\bk)}-n_j\pi_{(\bi\bk)})h^{\bk\a} \nn\\
 &&-\nabla_\b\bigl[(N^\a\phi_{ij}{^\b}-N^\b\phi_{ij}{^\a})\bigr]
  +\fr{1}{2}N(n_i\phi_{\bj\bk}-n_j\phi_{\bi\bk})h^{\bk\a} \, .\lab{C2} 
\eea

\subsub{2.} In order to calculate the Poisson brackets between $F_{ij}$
and the Hamiltonian constraints, we also need the following relations:
\bsubeq 
\bea
&&\{ B_{ij}^{o\a},\cH'_{kl} \}
   = (\eta_{ik}B_{lj}^{0\a}+ \eta_{jk}B_{il}^{0\a})\d -(kl)\, ,\nn\\
&&\{ B_{ij}^{o\a}, \cH'_\b \} =\nabla_\b( B_{ij}^{0\a}\d )
     +\d^\a_\b B_{ij}^{0\g}\pd'_\g \d\, ,                    \lab{C3a}
\eea
and
\be
\{ B_{ij}^{o\a},\cH'_\ort \}= {1\over a}
        \bigl( n_i\pi_{(\bj\bk)}-n_j\pi_{(\bi\bk)}\bigr)h^{k\a}\d
  +2\ve^{0\a\b\g}_{ijmn}b^m_\b\nabla_\g(n^n\d) \, .          \lab{C3b} 
\ee
\esubeq 

Combining $(C1a)$ and $(C3a)$ we find
\bsubeq 
\bea
\{ \Pi_{ij}{^\a}, \cH'_{kl} \}\,&&=
 \left(\eta_{ik}\Pi_{lj}{^\a} 
                   + \eta_{jk}\Pi_{il}{^\a}\right)\d -(kl)\, ,\nn\\
\{ \Pi_{ij}{^\a},\cH'_\b\}\,&&= -\d^\a_\b\phi_{ij}\d 
                   -\d^\a_\b\nabla_\g(\Pi_{ij}{^\g}\d)
                   +\nabla_\b(\Pi_{ij}{^\a}\d)\, ,          \lab{C4a}
\eea
which implies \eq{4.10a}.

Similarly, combining  $(C1b)$ and $(C3b)$, and using the identity
$$
B_{ij}^{\a\b}=2N\ve_{ijmn}^{\a\b\g 0}b^m_\g n^n 
               -(N^\a B_{ij}^{0\b}-N^\b B_{ij}^{0\a}) \, ,
$$
we obtain
\bea
&&\{\Pi_{ij}{^\a},\cH'_\ort\}= 
   \fr{1}{2}\bigl(n_i\phi_{\bj\bk}-n_j\phi_{\bi\bk}\bigr)h^{\bk\a}\d
   +\nabla_\b(M_{ij}^{\a\b}\d) \, ,\nn\\
&& NM_{ij}^{\a\b}\equiv  2aH^{\a\b}_{ij} +4b\l_{ij}{}^{\a\b} 
  +N^\a\bigl( \Pi_{ij}{}^{\b}-\phi_{ij}{^\b}\bigr)
  -N^\b\bigl( \Pi_{ij}{}^{\a}-\phi_{ij}{^\a}\bigr)  \, ,   \lab{C4b}
\eea
\esubeq 
which implies \eq{4.10b}.

\section{Extra gauge generators} 

In this Appendix we derive the form of the gauge generator \eq{6.5}.
We start with $G^\one_{ij}=\phi_{ij}$ in \eq{6.1b}. In order to find the
form of the accompanying component $G^\nul_{ij}$, we use the PB algebra
given in Eq. \eq{5.2}, and calculate 
$$
\{ \phi_{ij},H_T \}=-\fr{1}{2}[\nabla_\a,\nabla_\b]K_{ij}^{\a\b}\, ,
\qquad K_{ij}^{\a\b}\equiv  2aH_{ij}^{\a\b} +4b\l_{ij}{}^{\a\b}\, ,
$$ 
where terms proportional to $\phi_{ij}{^\a}$ are discarded.
The second condition in \eq{6.1b} implies
\bsubeq 
\be
G^\nul_{ij}=\fr{1}{2}\bigl[ R_i{^s}{}_{\a\b} K_{sj}^{\a\b} 
            -(ij)\bigr] +\th_{ij}\, ,\qquad             
\ee
where $\th_{ij}$ is a primary FC constraint. The third condition in
\eq{6.1b} can be written in the form 
$$
\fr{1}{2}\bigl[ \dot R_i{^s}{}_{\a\b} K_{sj}^{\a\b} 
 +R_i{^s}{}_{\a\b}\dot K_{sj}^{\a\b}-(ij)\bigr] +\dot\th_{ij}=C_{PFC} \, ,
$$
where $\dot X=\{ X,H_T \}$. Then, using the relations 
$\nabla_0 R_i{^s}{}_{\a\b}\approx 0$ and 
$\dot\pi^{is}{}_{\a\b}=4bR^{is}{}_{\a\b}$ we obtain
\be
\th_{ij}= {1\over 2}{1\over 4b}\left[  
 \left( A_i{^n}{_0} \pi_n{^s}{}_{\a\b} 
   +A^s{}_{n0}\pi_i{^n}{}_{\a\b} \right) K_{sj}^{\a\b} 
   -\pi_i{^s}{}_{\a\b}\dot K_{sj}^{\a\b}\right] -(ij) \, . 
\ee
\esubeq 

\end{document}